\documentstyle[aps,prd,eqsecnum,twoside,epsfig]{revtex}
\newcommand {\bx}{\mbox{\boldmath $\xi $}}
\newcommand {\ve}{\varepsilon}
\newcommand {\pr}{\partial}
\newcommand {\cG}{\cal G}
\newcommand {\cD}{\cal D}
\newcommand {\bg}{\bar \gamma}
\newcommand {\G}{\Gamma}
\newcommand {\bp}{\bar \psi}

\newcommand {\vf}{\varphi}
\newcommand {\vt}{\vartheta}
\begin{document}
\baselineskip -24pt
\title{Spinor Field in Bianchi type-I Universe: regular solutions} 
\author{Bijan Saha\\ 
Laboratory of Information Technologies\\ 
Joint Institute for Nuclear Research, Dubna\\ 
141980 Dubna, Moscow region, Russia\\ 
e-mail:  saha@thsun1.jinr.ru, bijan@cv.jinr.ru}
\maketitle 

\begin{abstract}
Self-consistent solutions to the nonlinear spinor field  equations 
in General Relativity has been studied for the case of Bianchi type-I (B-I)
space-time. It has been shown that, for some special type of nonliearity
the model provides regular solution, but this singularity-free solutions
are attained at the cost of broken dominant energy condition in
Hawking-Penrose theorem. It has also been shown that the introduction of 
$\Lambda$-term in the Lagrangian generates oscillations of the B-I model,
which is not the case in absence of $\Lambda$ term. Moreover, for the
linear spinor field, the $\Lambda$ term provides oscillatory solutions,
those are regular everywhere, without violating dominant energy condition.
\end{abstract}

\section{Introduction}
Nonlinear phenomena have been one of the most popular topics during 
last years. Nevertheless, it must be admitted that nonlinear classical 
fields have not received general consideration. This is probably due
to the mathematical difficulties which arise because of the 
nonrenormalizability of the Fermi and other nonlinear couplings~\cite{ranada}.
Nonlinear selfcouplings of the spinor fields may arise as a consequence 
of the geometrical structure of the space-time and, more precisely, 
because of the existence of torsion. As soon as 1938, Ivanenko
\cite{ivanenko1,ivanenko2,rodichev}
showed that a relativistic theory imposes in some cases a fourth order
selfcoupling. In 1950 Weyl~\cite{weyl} 
proved that, if the affine and the metric properties of the space-time
are taken as independent, the spinor field obeys either a linear equation
in a space with torsion or a nonlinear one in a Reimannian space. As
the selfaction is of spin-spin type, it allows the assignment of a 
dynamical role to the spin and offers a clue about the origin of the
nonlinearities. This question was further clarifed in some important
papers by Utiyama, Kibble and Sciama ~\cite{utiyama,kibble,sciama}
In the simplest scheme the selfaction is of pseudovector
type, but it can be shown that one can also get a scalar coupling
~\cite{soler}. An exelent review of the problem may be found in~\cite{hehl}. 
Nonlinear quantum Dirac fields were used by Heisenberg~\cite{hb1,hb2}
in his ambitious unified theory of elementary particles. They are 
presently the object of renewed interest since the widely known paper
by Gross and Neveu~\cite{gross} 

The quantum field theory in curved space-time has been a matter of
great interest in recent years because of its applications to 
cosmology and astrophysics. The evidence of existence of strong
gravitational fields in our Universe led to the study of the quantum 
effects of material fields in external classical gravitational field.
After the appearance of Parker's paper on scalar fields~\cite{parker1} and 
spin-$\frac{1}{2}$ fields,~\cite{parker2} several authors have studied this 
subject. The present cosmology is based largely on Friedmann's solutions
of the Einstein equations which describe the completely uniform and
isotropic universe ('closed' and 'open' models, i.e., bounded or
unbounded universe). The main feature of these solutions is their
non-stationarity. The idea of expanding universe, following from this
property, is confirmed by the astronomical observations and it now safe
to assume that the isotropic model provides, in its general features,
an adequate description of the present state of the universe.
Although the Universe seems homogenous and isotropic at present,
it does not necessarily mean that it is also suitable for description of 
the early stages of the development  of the universe and 
there are no observational data guarantying the isotropy in the era prior
to the recombination. In fact, there are theoretical arguments that sustain 
the existence of an anisotropic phase that approaches an isotropic one
~\cite{misner}. Interest in studying Klein-Gordon and Dirac equations in 
anisotropic models has increased since Hu and Parker~\cite{hu1} have 
shown that the creation of scalar particles in anisotropic backgrounds can
dissipate the anisotropy as the Universe expands. 

A Bianchi type-I (B-I) Universe, being the straightforward generalization 
of the flat Robertson-Walker (RW) Universe, is one of the simplest models  
of an anisotropic Universe that describes a homogenous and spatially flat
Universe. Unlike the RW Universe which has the same 
scale factor for each of the three spatial directions, a B-I Universe
has a different scale factor in each direction, thereby introducing an
anisotropy to the system. It moreover has the agreeable property that
near the singularity it behaves like a Kasner Universe, even in the 
presence of matter, and consequently falls within the general analysis
of the singularity given by Belinskii et al~\cite{belinskii}. 
Also in a Universe filled with matter for $p\,=\,\zeta\,\ve, \quad 
\zeta < 1$, it has been shown that any initial anisotropy in a B-I
Universe quickly dies away and a B-I Universe eventually evolves
into a FRW Universe~\cite{jacobs}. Since the present-day Universe is 
surprisingly isotropic, this feature of the B-I Universe makes it a prime 
candidate for studying the possible effects of an anisotropy in the early 
Universe on present-day observations. 
In light of the importance of mentioned 
above, several authors have studied B-I universe from different aspects.

In~\cite{chimento} Chimento and Mollerach studied the Dirac equations
in B-I universe and obtained their classical solutions. They also
claimed that for each value of the momentum only two independent 
solutions exist and showed that it is not possible to obtain the
solutions from those of a FRW universe only by perturbation. 
One of the solutions obtained would describe a particle with a 
given helicity, while the other one would represent antiparticles 
with the opposite helicity. This fact posed a very interesting problem, 
spin-$1/2$ particles cannot live in a B-I, at least if they keep their
well-known properties of flat space-time. This problem was handled
by Castagnino et al~\cite{castagnino} where they showed that if the 
Dirac equation is separable, the number of independent solution
is four, contrary to the claim made in ~\cite{chimento}.

In a number of papers~\cite{hu2,miedema,cho} several authors studied
the behavior of gravitational waves (GW's) in a B-I Universe. In
~\cite{miedema} the evolution equations for small perturbations in the
metric, energy density, and material velocity were derived for an
anisotropic viscous B-I universe.It has been shown that the results
were independent of the equation of state of the cosmic fluid and its 
viscosity. They also showed that the GWs need not 
necessarily be transversal in an anisotropically expanding B-I
universe and the longitudinal components of the gravitational waves
have no physical significance. In~\cite{cho} Cho and Speliotopoulos
studied the propagation of classical gravitational waves in B-I universe.
They found that GWs in B-I universe are not equivalent to two 
minimally coupled massless scalar fields as in FRW universe. Because
of its tensorial nature, the GW is much more sensitive to the anisotropy
in space-time than the scalar field is and it gains an effective mass term. 
Moreover, they found a coupling between the two polarization states
of the GW which is not present in a FRW universe.

Nonlinear spinor field (NLSF) in external FRW cosmological
gravitational field was first studied by G.N. Shikin in 
1991~\cite{shikin}. The main purpose to introduce a nonlinear
term in the spinor field Lagrangian is to study the possibility
of elimination of initial singularity. Following~\cite{shikin},
we analyzed the nonlinear spinor field equations in an external
B-I universe~\cite{sahapfu1}. In that paper we consider the nonlinear 
term in the spinor field Lagrangian as an arbitrary function of all 
possible invariants generated from spinor bilinear forms. There we also
studied the possibility of elimination of initial singularity especially 
for the Kasner Universe. For few years we studied the behavior of 
self-consistent NLSF in a B-I Universe~\cite{sahactp1,sahajmp} 
both in presence of perfect fluid and without it that was 
followed by the Refs.,~\cite{sahactp2,sahaizv,sahagrg} where we studied
the self-consistent system of interacting spinor and scalar fields.
Recently, we study~\cite{pfu-l,sahal} the role of the cosmological 
constant ($\Lambda$) in the Lagrangian which together with 
Newton's gravitational constant ($G$) is considered as the fundamental
constants in Einstein's theory of gravity~\cite{einstein}.

\section{Review of B-I cosmology}

A diagonal Bianchi type-I space-time (hereafter B-I)
is a spatially homogeneous space-time which admits an abelian group 
$G_3$, acting on spacelike hypersurfaces, generated by the spacelike 
Killing vectors $\bx_1 = \pr_{1},\, \bx_2 = \pr_{2},\,\bx_3 = \pr_{3}$. 
In synchronous coordinates the metric is \cite{bianchi,tsam}:
\begin{equation} 
ds^2 = dt^2 - \sum_{i=1}^{3}a_i^2 (t) dx_i^2. 
\label{BIb}
\end{equation}
If the three scale factors are equal (i.e., $ a_1 = a_2 = a_3$), 
(\ref{BIb}) describes an isotropic and spatially flat 
Friedmann-Robertson-Walker (FRW) universe. The B-I Universe
has a different scale factor in each direction, thereby introducing an
anisotropy to the system. Thus, a Bianchi type-I (B-I) universe, being 
the straightforward generalization of the flat Friedmann-Robertson-Walker 
(FRW) universe, is one of the simplest models of an anisotropic universe 
that describes a homogenous and spatially flat universe.  
When the two of the metric functions are equal (e.g. $a_2 = a_3$) 
the B-I space-time is reduce to the important class
of plane symmetric space-time (a special class of the Locally Rotational
Symmetric space-times~\cite{ellis1,ellis2}) which admit a $G_4$ group of 
isometries acting multiply transitively on the spacelike hypersurfaces 
of homogeneity generated by the vectors $\bx_1,\,\bx_2,\,\bx_3$ and
$\bx_4 = x^2\pr_{3} - x^3\pr_{2}$. 
The B-I has the agreeable property that near the singularity it behaves 
like a Kasner universe, given by
\begin{equation}
a_1(t) = a_{1}^{0} t^{p_1}, \quad
a_2(t) = a_{2}^{0} t^{p_2}, \quad
a_3(t) = a_{3}^{0} t^{p_3}, 
\label{kasner}
\end{equation}
with $p_j$ being the parameters of the B-I space-time which measure
the relative anisotropy between any two asymmetry axis and satisfy
the constraints
\begin{mathletters}
\label{kc}
\begin{eqnarray}
p_1 + p_2 + p_3 &=& 1 \\
p_{1}^{2} + p_{2}^{2} + p_{3}^{2} &=& 1. 
\end{eqnarray}
\end{mathletters}
Thus out of three parameters, only one is arbitrary. One particular
choice of parametrization is 
\begin{mathletters}
\label{kp1}
\begin{eqnarray}
p_{1} &=& \frac{-p}{p^2 + p +1} \\
p_{2} &=& \frac{p(p + 1)}{p^2 + p +1}\\
p_{3} &=& \frac{p+1}{p^2 + p +1}.
\end{eqnarray}
\end{mathletters}
The condition $0 \le p \le 1$ on $p$ then yields the condition
$-1/3 \le p_1 \le 0, \quad 0 \le p_2 \le 2/3, \quad 2/3 \le p_3 \le 1.$
Another particular parametrization can be given using an angle on the 
unit circle, since (\ref{kc}) describes the intersection of a sphere
with aplane in the parameter space $(p_1,p_2,p_3)$:
\begin{mathletters}
\label{kp2}
\begin{eqnarray}
p_{1} &=& \frac{1}{3}(1 + {\rm cos} \vt + \sqrt{3}{\rm sin} \vt), \\
p_{2} &=& \frac{1}{3}(1 + {\rm cos} \vt - \sqrt{3}{\rm sin} \vt),\\
p_{3} &=& \frac{1}{3}(1 - 2 {\rm cos} \vt). 
\end{eqnarray}
\end{mathletters}
Although $\vt$ ranges over the unit circle, the labeling of each $p_j$
is quite arbitraey. Thus the unit circle can be devided into six equal 
parts each of which span $60^o$, and the choice of $p_j$ is unique
within each section separately. For $\vt =0$, $p_1 = p_2 = 2/3$ and
$p_3 = -1/3$ while for $\vt = \pi/3$ $p_1 = 1$ and $p_2 = p_3 = 0$.

Let us now go back to B-I metric. The non-trivial Christoffel
symbols for (\ref{BIb}) are
\begin{eqnarray}
\G_{ii}^{0} = a_i {\dot a_i}, \quad
\G_{0i}^{i} = \G_{i0}^{i} = \frac{\dot a_i}{a_i}, 
\label{chris}
\end{eqnarray}
while the components of non-trivial Ricci tensor read
\begin{eqnarray}
R_{00} = - \sum_{i=1}^{3} \frac{\ddot a_i}{a_i}, \quad 
R_{ii} = \Bigl[\frac{\ddot a_i}{a_i} + \frac{\dot a_i}{a_i}
\Bigl(\frac{\dot a_j}{a_j} + \frac{\dot a_k}{a_k}\Bigr)\Bigr] a_i^2,
\quad i,j,k = 1,2,3, \quad i \ne j \ne k. 
\label{RT}
\end{eqnarray}
The Ricci scalar for the B-I universe has the form
\begin{eqnarray}
 R = -2 \Bigl(\frac{\ddot a_1}{a_1} + \frac{\ddot a_2}{a_2} + 
\frac{\ddot a_3}{a_3}
+\frac{\dot a_1}{a_1}\frac{\dot a_2}{a_2} + \frac{\dot a_2}{a_2}
\frac{\dot a_3}{a_3} + \frac{\dot a_3}{a_3}\frac{\dot a-1}{a_1}\Bigr).
\label{RS}
\end{eqnarray}
 
Sometimes it proves convenient to introduce a new time parameter
$\eta$ by
\begin{equation}
\eta = \int\limits_{}^{t} a^{-1} (\bar t) d{\bar t},
\label{newtime}
\end{equation}
where we define
\begin{equation}
[a(t)]^2 = C(t) \equiv (a_1 a_2 a_3)^{2/3} = (C_1 C_2 C_3)^{1/3}, 
\end{equation}
with $C-i \equiv a_{i}^{2}$. Note that in the isotropic limit, i.e.,
$a_1 = a_2 = a_3$ $\eta$ reduces to conformal time. Further, defining
\begin{equation}
d_i = \frac{C_i^{\prime}}{C_i}, \quad 
D \equiv \frac{1}{3}\sum_{i=1}^{3}d_i = \frac{C^{\prime}}{C}, \quad
Q \equiv \frac{1}{72}\sum_{i < j} (d_i - d_j)^2
\end{equation} 
where prime ($\prime$) denotes differentiation with respect to $\eta$,
we get the following nonzero Christoffel symbols for the metric (\ref{BIb})
\begin{equation}
\G_{\eta \eta}^{\eta} = \frac{1}{2}D, \quad
\G_{ii}^{\eta} = \frac{1}{2}\frac{d_i C_i}{C}, \quad
\G_{i \eta}^{i}=  \G_{\eta i}^{i} = \frac{1}{2} d_i.
\end{equation}
The nonzero components of the Ricci tensor now read
\begin{equation}
R_{\eta \eta} = \frac{3}{2} D^{\prime} + 6 Q, \quad
R_{ii} = - \frac{C_i}{2 C} (d_{i}^{\prime} + d_i D)
\end{equation} 
and the Ricci scalar
\begin{equation}
R = C^{-1} [ 3 D^{\prime} + \frac{3}{2} D^2  + 6 Q].
\end{equation}
Note the in the sections to follow, we work with the usual time $t$.

\section{Fundamental Equations and general solutions}

The action of the nonlinear spinor, scalar and gravitational fields
can be written as
\begin{equation}
{\cal S}(g; \psi, \bp, \vf) = \int\, L \sqrt{-g} d\Omega
\label{action}
\end{equation}
with 
\begin{equation} 
L= L_{\rm g} + L_{\rm sp} + L_{\rm m}.
\label{lag} 
\end{equation} 
Here $L_{\rm g}$ corresponds to the gravitational field 
\begin{equation}
L_{\rm g} = \frac{R + 2 \Lambda}{2\kappa},
\label{lgrav}
\end{equation}
where $R$ is the scalar curvature, $\kappa = 8 \pi G$ with G 
being the Einstein's gravitational constant and $\Lambda$ is 
the cosmological constant. The spinor field Lagrangian $L_{\rm sp}$ 
is given by
\begin{equation}
L_{\rm sp} = \frac{i}{2} 
\biggl[\bp \gamma^{\mu} \nabla_{\mu} \psi- \nabla_{\mu} \bar 
\psi \gamma^{\mu} \psi \biggr] - m\bp \psi + L_{\rm N},
\label{lspin}
\end{equation}
where the nonlinear term $L_{\rm N}$ describes the self-interaction of a 
spinor field and can be presented as some arbitrary functions of 
invariants generated from the real bilinear forms of a spinor 
field. Since $\psi$ and $\psi^{\star}$ (complex conjugate of $\psi$)
has $4$ component each, one can construct $4\cdot 4 = 16$ independent
bilinear combinations. They are
\begin{mathletters}
\label{bf}
\begin{eqnarray}
 S&=& \bar \psi \psi,\qquad \qquad \,\,\,\,({\rm scalar}),   \\                   
  P&=& i \bar \psi \gamma^5 \psi,\qquad \quad \,\,({\rm pseudoscalar}), \\
 v^\mu &=& (\bar \psi \gamma^\mu \psi), \qquad \quad ({\rm vector})\\
 A^\mu &=&(\bar \psi \gamma^5 \gamma^\mu \psi),\qquad\, ({\rm pseudovector}), \\
T^{\mu\nu} &=&(\bar \psi \sigma^{\mu\nu} \psi),\,\,\,\,\qquad
({\rm antisymmetric\,\,\, tensor}),  
\end{eqnarray}
\end{mathletters}
where $\sigma^{\mu\nu}\,=\,(i/2)[\gamma^\mu\gamma^\nu\,-\,
\gamma^\nu\gamma^\mu]$. 
Invariants, corresponding to the bilnear forms are
\begin{mathletters}
\label{invariants}
\begin{eqnarray}
I &=& S^2, \\
J &=& P^2, \\ 
I_v &=& v_\mu\,v^\mu\,=\,(\bar \psi \gamma^\mu \psi)\,g_{\mu\nu}
(\bar \psi \gamma^\nu \psi),\\ 
I_A &=& A_\mu\,A^\mu\,=\,(\bar \psi \gamma^5 \gamma^\mu \psi)\,g_{\mu\nu}
(\bar \psi \gamma^5 \gamma^\nu \psi), \\
I_T &=& T_{\mu\nu}\,T^{\mu\nu}\,=\,(\bar \psi \sigma^{\mu\nu} \psi)\,
g_{\mu\alpha}g_{\nu\beta}(\bar \psi \sigma^{\alpha\beta} \psi). 
\end{eqnarray}
\end{mathletters}

According to the Pauli-Fierz theorem,\cite{landau} among the five invariants
only $I$ and $J$ are independent as all other can be expressed by them:
$I_v = - I_A = I + J$ and $I_T = I - J.$ Therefore we choose the nonlinear
term $F$ to be the function of $I$ and $J$ only, i.e., $L_{\rm N} = F(I, J)$, 
thus claiming that it describes the nonlinearity in the most general of 
its form. 
$L_{\rm m}$ is the Lagrangian of perfect fluid. 

Variation of (\ref{action}) with respect to spinor field $\psi\,(\bp)$
gives nonlinear spinor field equations
\begin{mathletters}
\label{speq}
\begin{eqnarray}
i\gamma^\mu \nabla_\mu \psi - m \psi + {\cD} \psi + 
{\cG} i \gamma^5 \psi &=&0, \label{speq1} \\
i \nabla_\mu \bp \gamma^\mu +  m \bp - {\cD} \bp - 
{\cG} i \bp \gamma^5 &=& 0, \label{speq2}
\end{eqnarray}
\end{mathletters}
where we denote
$$ {\cD} = 2 S \frac{\pr F}{\pr I}, \quad
{\cG} = 2 P \frac{\pr F}{\pr J}.$$

Varying (\ref{action}) with respect to metric tensor $g_{\mu\nu}$ 
one finds the Einstein's field equation
\begin{equation}
R_{\nu}^{\mu} - \frac{1}{2}\,\delta_{\nu}^{\mu} R = - \kappa 
T_{\nu}^{\mu} + \Lambda \delta_{\nu}^{\mu} 
\label{ee}
\end{equation} 
where $R_{\nu}^{\mu}$ is the Ricci tensor; $R = g^{\mu\,\nu} R_{\mu\,\nu}$
is the Ricci scalar; and $T_{\nu}^{\mu}$ is the energy-momentum tensor
of the material field given by
\begin{equation}
T_{\mu}^{\nu} = T_{{\rm sp}\,\mu}^{\,\,\,\nu} + 
T_{{\rm m}\,\mu}^{\,\,\,\nu}.
\label{tem}
\end{equation}
Here $T_{{\rm sp}\,\mu}^{\,\,\,\nu}$ is the energy-momentum tensor of 
the spinor field  
\begin{equation}
T_{{\rm sp}\,\mu}^{\,\,\,\rho}=\frac{i}{4} g^{\rho\nu} \biggl(\bp \gamma_\mu 
\nabla_\nu \psi + \bp \gamma_\nu \nabla_\mu \psi - \nabla_\mu \bar 
\psi \gamma_\nu \psi - \nabla_\nu \bp \gamma_\mu \psi \biggr) \,-
\delta_{\mu}^{\rho}L_{sp}
\label{temsp}
\end{equation}
where $L_{sp}$ with respect to (\ref{speq}) takes the form
\begin{equation}
L_{sp} = -\bigl({\cD} S + {\cG} P\bigr) + F(I,J).
\label{lsp}
\end{equation}
$T_{\mu\,(m)}^{\nu}$ is the energy-momentum tensor of a perfect fluid. 
For a Universe filled with perfect fluid, in the concomitant system of 
reference $(u^0=1, \, u^i=0, i=1,2,3)$ we have
\begin{equation}
T_{\mu (m)}^{\nu}\,=\, (p + \ve) u_\mu u^\nu - 
\delta_{\mu}^{\nu} p \,=\,(\ve,\,- p,\,- p,\,- p),
\end{equation} 
where energy $\ve$ is related to the pressure $p$ by the equation 
of state $p\,=\,\zeta\,\ve$. The general solution has been derived 
by Jacobs~\cite{jacobs}. Here $\zeta$ varies between the
interval $0\,\le\, \zeta\,\le\,1$, whereas $\zeta\,=\,0$ describes
the dust Universe, $\zeta\,=\,\frac{1}{3}$ presents radiation Universe,
$\frac{1}{3}\,<\,\zeta\,<\,1$ ascribes hard Universe and $\zeta\,=\,1$
corresponds to the stiff matter. 

In (\ref{speq}) and (\ref{tem}) 
$\nabla_\mu$ denotes the covariant differentiation; its explicit form 
depends on the quantity it acts on. This covariant differentiation
has the standard properties
\begin{mathletters}
\label{nabla}
\begin{eqnarray}
\nabla_\mu (AB) &=& (\nabla_\mu A) B + A (\nabla_\mu B),\\
\nabla_\mu (A^*) &=& (\nabla_\mu A)^*,\\
\nabla_\mu \gamma_\nu &=& 0, \label{nablag}
\end{eqnarray}  
\end{mathletters}
where the symbol $^*$ means Hermitian adjoint (the transpose of the 
complex conjugate). The explicit form of the covariant derivative
of spinor is ~\cite{zhelnorovich,brill}
\begin{mathletters}
\label{cvd}
\begin{eqnarray} 
\nabla_\mu \psi &=& \frac{\partial \psi}{\partial x^\mu} -\G_\mu \psi, \\
\nabla_\mu \bp &=& \frac{\partial \bp}{\partial x^\mu} + \bp \G_\mu, 
\end{eqnarray} 
\end{mathletters}
where $\G_\mu(x)$ are spinor affine connection matrices. 
$\gamma$ matrices in the above equations obey the following algebra
\begin{equation}
\gamma^\mu \gamma^\nu + \gamma^\nu \gamma^\mu = 2 g^{\mu\nu}
\label{al}
\end{equation}
and are connected with 
the flat space-time Dirac matrices $\bg$ in the following way
\begin{equation}
 g_{\mu \nu} (x)= e_{\mu}^{a}(x) e_{\nu}^{b}(x) \eta_{ab}, 
\quad \gamma_\mu(x)= e_{\mu}^{a}(x) \bg_a, 
\label{dg}
\end{equation}
where $\eta_{ab}= {\rm diag}(1,-1,-1,-1)$ and $e_{\mu}^{a}$ is a 
set of tetrad 4-vectors. The spinor affine connection matrices 
$\G_\mu (x)$ are uniquely determined up to an additive multiple
of the unit matrix by the equation
\begin{equation}
\nabla_\mu \gamma_\nu = \frac{\pr \gamma_\nu}{\pr x^\mu}
- \G_{\nu\mu}^{\rho}\gamma_\rho - \G_\mu \gamma_\nu
+ \gamma_\nu \G_\mu = 0,
\label{afsp}
\end{equation}
with the solution 
\begin{equation}
\G_\mu (x)= 
\frac{1}{4}g_{\rho\sigma}(x)\biggl(\partial_\mu e_{\delta}^{b}e_{b}^{\rho} 
- \G_{\mu\delta}^{\rho}\biggr)\gamma^\sigma\gamma^\delta. 
\label{gm}
\end{equation}
Let us now write the $\gamma$'s and $\G_\mu$'s explicitly for the
B-I metric (\ref{BIb}) that we rewrite in the form ~\cite{zeldovich}
\begin{equation} 
ds^2 = dt^2 - a^2(t) dx^2 - b^2(t) dy^2 - c^2(t) dz^2. 
\label{BI1}
\end{equation}
For the metric (\ref{BI1}) from (\ref{dg}) one finds
\begin{eqnarray}
\gamma_0 &=& \bg_0,\quad \gamma_1 = a(t)\bg_1,\quad 
\gamma_2= b(t)\bg_2,\quad \gamma_3 = c(t) \bg_3, \nonumber\\ \\
\gamma^0 &=& \bg^0,\quad \gamma^1 =\bg^1 /a(t),\quad 
\gamma^2= \bg^2 /b(t),\quad \gamma^3 = \bg^3 /c(t). 
\nonumber
\end{eqnarray}
For the affine spinor connections from (\ref{gm}) we find
\begin{eqnarray} 
\G_0 = 0, \quad 
\G_1 = \frac{1}{2}\dot a(t) \bg^1 \bg^0, \quad
\G_2 = \frac{1}{2}\dot b(t) \bg^2 \bg^0, \quad 
\G_3 = \frac{1}{2}\dot c(t) \bg^3 \bg^0. 
\end{eqnarray}
Flat space-time matrices $\bg$ we will choose in the form, 
given in~\cite{bogoliubov}:
\begin{eqnarray}
\bg^0&=&\left(\begin{array}{cccc}1&0&0&0\\0&1&0&0\\
0&0&-1&0\\0&0&0&-1\end{array}\right), \quad
\bg^1\,=\,\left(\begin{array}{cccc}0&0&0&1\\0&0&1&0\\
0&-1&0&0\\-1&0&0&0\end{array}\right), \nonumber\\
\bg^2&=&\left(\begin{array}{cccc}0&0&0&-i\\0&0&i&0\\
0&i&0&0\\-i&0&0&0\end{array}\right), \quad
\bg^3\,=\,\left(\begin{array}{cccc}0&0&1&0\\0&0&0&-1\\
-1&0&0&0\\0&1&0&0\end{array}\right).  \nonumber
\end{eqnarray}
Defining $\gamma^5$ as follows,
\begin{eqnarray}
\gamma^5&=&-\frac{i}{4} E_{\mu\nu\sigma\rho}\gamma^\mu\gamma^\nu
\gamma^\sigma\gamma^\rho, \quad E_{\mu\nu\sigma\rho}= \sqrt{-g}
\ve_{\mu\nu\sigma\rho}, \quad \ve_{0123}=1,\nonumber \\
\gamma^5&=&-i\sqrt{-g} \gamma^0 \gamma^1 \gamma^2 \gamma^3 
\,=\,-i\bg^0\bg^1\bg^2\bg^3 =
\bg^5, \nonumber
\end{eqnarray}
we obtain
\begin{eqnarray}
\bg^5&=&\left(\begin{array}{cccc}0&0&-1&0\\0&0&0&-1\\
-1&0&0&0\\0&-1&0&0\end{array}\right).\nonumber
\end{eqnarray}
For the space-time (\ref{BI1}) Einstein equations (\ref{ee}) now read
\begin{mathletters}
\label{BID}
\begin{eqnarray}
\frac{\ddot b}{b} +\frac{\ddot c}{c} + \frac{\dot b}{b}\frac{\dot 
c}{c}&=&  \kappa T_{1}^{1} -\Lambda,\label{11}\\
\frac{\ddot c}{c} +\frac{\ddot a}{a} + \frac{\dot c}{c}\frac{\dot 
a}{a}&=&  \kappa T_{2}^{2} - \Lambda,\label{22}\\
\frac{\ddot a}{a} +\frac{\ddot b}{b} + \frac{\dot a}{a}\frac{\dot 
b}{b}&=&  \kappa T_{3}^{3} - \Lambda,\label{33}\\
\frac{\dot a}{a}\frac{\dot b}{b} +\frac{\dot b}{b}\frac{\dot c}{c} 
+\frac{\dot c}{c}\frac{\dot a}{a}&=&  \kappa T_{0}^{0} - \Lambda,
\label{00}
\end{eqnarray}
\end{mathletters}
where point means differentiation with respect to t. 
 
We will study the space-independent solutions to the spinor 
field equations (\ref{speq}) so that 
$\psi=\psi(t)$.
Setting
\begin{equation}
\tau = a b c = \sqrt{-g}
\label{taudef}
\end{equation}
we rewrite the spinor field equation (\ref{speq1}) as
\begin{equation} i\bg^0 
\biggl(\frac{\partial}{\partial t} +\frac{\dot \tau}{2 \tau} \biggr) \psi 
- m \psi + {\cD}\psi + {\cG} i \gamma^5 \psi = 0.
\label{spq}
\end{equation} 
Setting $V_j(t) = \sqrt{\tau} \psi_j(t), \quad j=1,2,3,4,$ from 
(\ref{spq}) one deduces the following system of equations:  
\begin{mathletters}
\label{V}
\begin{eqnarray} 
\dot{V}_{1} + i (m - {\cD}) V_{1} - {\cG} V_{3} &=& 0, \\
\dot{V}_{2} + i (m - {\cD}) V_{2} - {\cG} V_{4} &=& 0, \\
\dot{V}_{3} - i (m - {\cD}) V_{3} + {\cG} V_{1} &=& 0, \\
\dot{V}_{4} - i (m - {\cD}) V_{4} + {\cG} V_{2} &=& 0. 
\end{eqnarray} 
\end{mathletters}
Using the solutions obtained one can write the components of
spinor current:
\begin{equation}
j^\mu = \bp \gamma^\mu \psi.
\label{spcur}
\end{equation}
Taking into account that $\bp = \psi^{\dagger} \bg^0$, where 
$\psi^{\dagger} = \bigl(\psi_{1}^{*},\,\psi_{2}^{*},\,\psi_{3}^{*},\,
\psi_{4}^{*}\bigr)$ and $\psi_j = V_j/\sqrt{\tau}, \quad j=1,2,3,4$
for the components of spin current we write
\begin{mathletters}
\label{spincur}
\begin{eqnarray}
j^0 &=& \frac{1}{\tau}
\bigl[V_{1}^{*} V_{1} + V_{2}^{*} V_{2} + V_{3}^{*} V_{3}
+ V_{4}^{*} V_{4}\bigr],\\
j^1 &=& \frac{1}{a\tau}
\bigl[V_{1}^{*} V_{4} + V_{2}^{*} V_{3} + V_{3}^{*} V_{2}
+ V_{4}^{*} V_{1}\bigr],\\
j^2 &=& \frac{-i}{b\tau}
\bigl[V_{1}^{*} V_{4} - V_{2}^{*} V_{3} + V_{3}^{*} V_{2}
- V_{4}^{*} V_{1}\bigr],\\
j^3 &=& \frac{1}{c\tau}
\bigl[V_{1}^{*} V_{3} - V_{2}^{*} V_{4} + V_{3}^{*} V_{1}
- V_{4}^{*} V_{2}\bigr].
\end{eqnarray}
\end{mathletters}
The component $j^0$ defines the charge density of spinor field 
that has the following chronometric-invariant form 
\begin{equation}
\varrho = (j_0\cdot j^0)^{1/2}. 
\label{rho}
\end{equation}
The total charge of spinor field is defined as
\begin{equation}
Q = \int\limits_{}^{} \varrho \sqrt{-^3 g} dx dy dz
\label{charge}
\end{equation}

Let us consider the spin tensor~\cite{bogoliubov}
\begin{equation}
S^{\mu\nu,\epsilon} = \frac{1}{4}\bp \bigl\{\gamma^\epsilon
\sigma^{\mu\nu}+\sigma^{\mu\nu}\gamma^\epsilon\bigr\} \psi.
\label{spin}
\end{equation}
We write the components $S^{ik,0}$ $(i,k=1,2,3)$, defining the spatial
density of spin vector explicitly. From (\ref{spin}) we have
\begin{equation}
S^{ij,0} = \frac{1}{4}\bp \bigl\{\gamma^0
\sigma^{ij}+\sigma^{ij}\gamma^0\bigr\} \psi = 
\frac{1}{2}\bp \gamma^0 \sigma^{ij}\psi
\label{spin0}
\end{equation}
that defines the projection of spin vector on $k$ axis. Here $i,\,j,\,k$
takes the value $1,\,2,\,3$ and $i\ne j\ne k$. Thus, for the projection 
of spin vectors on the $X,\,Y$ and $Z$ axis we find
\begin{mathletters}
\label{spinvec}
\begin{eqnarray}
S^{23,0} &=& \frac{1}{2bc\tau}
\bigl[V_{1}^{*} V_{2} + V_{2}^{*} V_{1} + V_{3}^{*} V_{4}
+ V_{4}^{*} V_{3}\bigr],\\
S^{31,0} &=& \frac{-i}{2ca\tau}
\bigl[V_{1}^{*} V_{2} - V_{2}^{*} V_{1} + V_{3}^{*} V_{4}
- V_{4}^{*} V_{3}\bigr],\\
S^{12,0} &=& \frac{1}{2ab\tau}
\bigl[V_{1}^{*} V_{1} - V_{2}^{*} V_{2} + V_{3}^{*} V_{3}
- V_{4}^{*} V_{4}\bigr].
\end{eqnarray}
\end{mathletters}

The chronometric invariant spin tensor takes the form
\begin{equation}
S_{{\rm ch}}^{ij,0} = \bigl(S_{ij,0} S^{ij,0}\bigr)^{1/2},
\label{chij}
\end{equation} 
and the projection of the spin vector on $k$ axis is defined by
\begin{equation}
S_k = \int\limits_{-\infty}^{\infty} S_{{\rm ch}}^{ij,0} 
\sqrt{-^3 g} dx dy dz. 
\label{proj}
\end{equation} 
From (\ref{speq}) we also write the equations for the invariants
$ S = \bp \psi, \quad P = i \bp \gamma^5 \psi$ and    
$A = \bp \bg^5 \bg^0 \psi$
\begin{mathletters}
\label{inv}
\begin{eqnarray}
{\dot S_0} - 2 {\cG}\, A_0 &=& 0, \label{S0}\\
{\dot P_0} - 2 (m - {\cD})\, A_0 &=& 0, \label{P0}\\
{\dot A_0} + 2 (m - {\cD})\, P_0 + 2 {\cG} S_0 &=& 0, \label{A0} 
\end{eqnarray}
\end{mathletters}
where $S_0 = \tau S, \quad P_0 = \tau P$, and $ A_0 = \tau A$,
leading to the following relation
\begin{equation}
S^2 + P^2 + A^2 =  C^2/ \tau^2, \qquad C^2 = {\rm const.}
\label{inv1}
\end{equation}

Let us now solve the Einstein equations. To do it we first write the 
expressions for the components of the energy-momentum tensor explicitly. 
Using the property of flat space-time Dirac matrices and the explicit 
form of covariant derivative $\nabla_\mu$, one can easily find
\begin{eqnarray}
\label{temc}
T_{0}^{0} = mS - F(I,J)  + \ve \nonumber\\ \\ 
T_{1}^{1}=T_{2}^{2}=T_{3}^{3}= {\cD} S + {\cG} P - F(I,J)- p. \nonumber 
\end{eqnarray}

Summation of Einstein equations (\ref{11}), (\ref{22}),(\ref{33}) and 
(\ref{00}) multiplied by 3 gives
\begin{equation}
\frac{\ddot 
\tau}{\tau}= \frac{3}{2}\kappa \Bigl(T_{1}^{1}+T_{0}^{0}\Bigr) - 3 \Lambda 
\label{dtau}
\end{equation} 
For the right-hand-side of (\ref{dtau}) to be a function
of $\tau$ only, the solution to this equation is well-known~\cite{kamke}.
As we see in the next section, the right-hand-side of (\ref{dtau}) is
indeed a function of $\tau$.
Given the explicit form of $L_{\rm N}$ and $L_{\rm int}$ from (\ref{dtau}) 
one finds the concrete solution for $\tau$ in quadrature. 

Let us express $a, b, c$ through $\tau$. For this we notice that
subtraction of Einstein equations  (\ref{22}) and (\ref{11})  leads  to  
the equation 
\begin{equation}
\frac{\ddot a}{a}-\frac{\ddot b}{b}+\frac{\dot a \dot c}{ac}- 
\frac{\dot b \dot c}{bc}= \frac{d}{dt}\biggl(\frac{\dot a}{a}- 
\frac{\dot b}{b}\biggr)+\biggl(\frac{\dot a}{a}- \frac{\dot b}{b} \biggr) 
\biggl (\frac{\dot a}{a}+\frac{\dot b}{b}+ \frac{\dot c}{c}\biggr)= 0. 
\end{equation} 
with the solution
\begin{equation}
\frac{a}{b}= D_1 \mbox{exp} \biggl(X_1 \int \frac{dt}{\tau}\biggr), \quad 
D_1=\mbox{const.}, \quad X_1= \mbox{const.} 
\label{ab}
\end{equation}
Analogically, one finds
\begin{equation} 
\frac{a}{c}= D_2 \mbox{exp} \biggl(X_2 \int \frac{dt}{\tau}\biggr), \quad 
\frac{b}{c}= D_3 \mbox{exp} \biggl(X_3 \int \frac{dt}{\tau}\biggr),  
\label{ac}
\end{equation}
where $D_2, D_3, X_2, X_3 $ are integration constants. In view of
(\ref{taudef}) we find the following functional dependence between the 
constants $D_1,\, D_2,\, D_3,\, X_1,\, X_2,\, X_3 $:  
$$ D_2=D_1\, D_3, \qquad X_2= X_1\,+\,X_3.$$
Finally, from (\ref{ab}) and (\ref{ac}) we write $a(t), b(t)$, and $c(t)$ 
in the explicit form  
\begin{mathletters}
\label{abc}
\begin{eqnarray} 
a(t) &=& 
(D_{1}^{2}D_{3})^{1/3}\tau^{1/3}\mbox{exp}\biggl[\frac{2 X_1 + X_3 
}{3} \int\,\frac{dt}{\tau (t)} \biggr], \label{a} \\
b(t) &=& 
(D_{1}^{-1}D_{3})^{1/3}\tau^{1/3}\mbox{exp}\biggl[-\frac{X_1 - X_3 
}{3} \int\,\frac{dt}{\tau (t)} \biggr], \label{b}\\
c(t) &=& 
(D_{1}D_{3}^{2})^{-1/3}\tau^{1/3}\mbox{exp}\biggl[-\frac{X_1 + 2 X_3 
}{3} \int\,\frac{dt}{\tau (t)} \biggr]. \label{c}
\end{eqnarray}
\end{mathletters}
Thus the system of Einstein's equations is completely integrated. 

Defining Hubble constant in analogy with a FRW universe
from (\ref{abc}) we obtain
\begin{equation}
H_{j} = \frac{\dot a_j}{a_j} = \frac{{\dot \tau} + Y_j}{3 \tau}, \quad
j=1,2,3,
\label{hcs}
\end{equation}
or a generalized one
\begin{equation}
H = (H_1 + H_2 + H_3)/3 = {\dot \tau}/3 \tau
\label{hc}
\end{equation}
Here $a_1 = a,\,\, a_2 = b, \,\, a_3 = c$.
The decelaration parameter given by
\begin{equation}
q = -\frac{{\ddot R} R}{{\dot R}^2}
\end{equation}
for a FRW universe with $R$ being the scale factor can also be generalized
for the B-I space-time to obtain 
\begin{eqnarray}
q_i = - \frac{{\ddot a_i} a_i}{{\dot a_i}^2} = - 
\Bigl[\Bigl(\frac{\ddot a_i}{a_i}\Bigr)/\Bigl(\frac{\dot a_i}{a_i}\Bigr)^2
\Bigr] = - \Bigl[1 + \Bigl(\frac{\dot a_i}{a_i}\Bigr)^{\cdot}/
\Bigl(\frac{\dot a_i}{a_i}\Bigr)^2\Bigr].
\label{dp}
\end{eqnarray}
Inserting (\ref{abc}) into (\ref{dp}) one obtains
\begin{equation}
q_i = - \frac{{\ddot \tau} - 2 {\dot \tau}^2 - Y_i {\dot \tau} + Y_i^2}
{{\dot \tau}^2 + 2 Y_i {\dot \tau} + Y_i^2},\qquad i =1,2,3. 
\end{equation} 

Let us now go back to the Einstein equation (\ref{ee}). 
Taking the divergence of Einstein equation we obtain
\begin{equation}
T_{\mu;\nu}^{\nu} = T_{\mu,\nu}^{\nu} + \G_{\rho\nu}^{\nu} T_{\mu}^{\rho}
- \G_{\mu\nu}^{\rho} T_{\rho}^{\nu} = 0
\end{equation}
which in our case reads
\begin{equation}
{\dot T}_{0}^{0} + \frac{\dot \tau}{\tau}\bigl(T_{0}^{0} - T_{1}^{1}\bigr)
= 0.
\label{conserv}
\end{equation}
Putting $T_{0}^{0}$ and $T_{1}^{1}$ into (\ref{conserv}) we obtain
\begin{eqnarray}
 {\dot \ve} + (\ve + p)\frac{\dot \tau}{\tau} +
(m - {\cD}){\dot S_0} -  {\cG}{\dot P_0}
= 0, 
\label{dott0}
\end{eqnarray}
where $S_0 = \tau S$ and $P_0 = \tau P$. From (\ref{S0}) and (\ref{P0})
we have $(m - {\cD}){\dot S}_0 - {\cG}{\dot P}_0 = 0$. Further taking
into account the equation of state, i.e., $p = \zeta \ve$ we find
\begin{equation}
\frac{d \ve}{(1 + \zeta) \ve} + \frac{d \tau}{\tau} = 0,
\end{equation}
with the solutions
\begin{equation}
\ve = \frac{\ve_0}{\tau^{1+\zeta}},\quad 
p = \frac{\zeta \ve_0}{\tau^{1+\zeta}}
\label{vep}
\end{equation}
where $\ve_0$ is the integration constant. Note that the relation
(\ref{vep}) holds for any combination of the material field
Lagrangian, e.g., spinor or scalar or interacting spinor and scalar
fields. Thus we see that
the right-hand side of (\ref{dtau}) is a function of $\tau$ only.
Then (\ref{dtau}), multiplied by $2 {\dot \tau}$ can be written as
\begin{equation}
2 {\dot \tau}\,{\ddot \tau} = \bigl[3\bigl(\kappa ( T_{1}^{1} + T_{0}^{0})
- 2 \Lambda \bigr) \tau\bigr] {\dot \tau} = \Psi(\tau) {\dot \tau}  
\label{taug}
\end{equation} 
Solution to the equation (\ref{taug}) we write in quadrature
\begin{equation}
\int\,\frac{d \tau}{\sqrt{\int \Psi (\tau) d \tau}} = t.
\label{quad}
\end{equation}
Given the explicit form of $F(I,\,J)$, from (\ref{quad}) one finds 
concrete function $\tau(t)$. Once the value of $\tau$ is obtained, one can 
get expressions for components $\psi_j(t), \quad j = 1, 2, 3, 4.$
Thus the initial systems of Einstein and Dirac equations have been 
completely integrated.

Further we will investigate the existence of singularity (singular point)
of the gravitational case, that can be done investigating the
invariant characteristics of the space-time. In general relativity 
these invariants are composed from the curvature tensor and the 
metric one. Contrary to the electrodynamics, where there are two 
invariants only ($J_1 = F_{\mu\nu} F^{\mu\nu}$ and 
$J_2 = {\star}F_{\mu\nu} F^{\mu\nu}$), in 4-D Riemann space-time there
are 14 independent invariants. They are~\cite{mitskevich}
\begin{mathletters}
\label{}
\begin{eqnarray}
I_1 &=& R \\
I_2 &=& R_{\mu\nu}R^{\mu\nu},\\
I_3 &=& R_{\alpha\beta\mu\nu}R^{\alpha\beta\mu\nu},\\
I_4 &=& {\star}R_{\alpha\beta\mu\nu}R^{\alpha\beta\mu\nu},\\
I_5 &=& R^{\alpha}_{\beta}R^{\beta}_{\mu}R^{\mu}_{\alpha},\\
I_6 &=& R^{\alpha\beta}R^{\mu\nu}R_{\alpha\mu\beta\nu},\\
I_7 &=& R^{\alpha\beta}R^{\mu\nu}{\star}R_{\alpha\mu\beta\nu},\\
I_8 &=& R^{\alpha\beta\mu\nu}R_{\alpha\beta\sigma\rho}
        R_{\,\,\,\,\,\,\mu\nu}^{\sigma\rho},\\
I_9 &=& {\star}R^{\alpha\beta\mu\nu}R_{\alpha\beta\sigma\rho} 
        R_{\,\,\,\,\,\,\mu\nu}^{\sigma\rho},\\
I_{10} &=& R_{\alpha}^{\beta}R^{\alpha\mu}R_{\mu\nu}R_{\beta}^{\nu},\\
I_{11} &=& R_{\nu}^{\mu}R_{\rho\mu}^{\,\,\,\,\,\,\sigma\alpha}
           R_{\sigma\alpha}^{\,\,\,\,\,\,\beta [ \nu}R_{\beta}^{\rho\,\,\,]},\\
I_{12} &=& R_{\nu}^{\mu}{\star}R_{\,\,\,\,\,\,\rho\mu}^{\sigma\alpha}
           R_{\sigma\alpha}^{\,\,\,\,\,\,\beta [ \nu}R_{\beta}^{\rho\,\,\,]},\\
I_{13} &=& R_{\,\,\,\,\,\,\alpha\beta}^{\mu\nu}\bigl(
           A_{\,\,\,\,\,\,\mu\nu}^{\alpha\beta} +
           R_{\rho}^{\alpha}R_{\sigma}^{\rho}R_{\eta}^{\sigma}
           R_{\mu}^{\eta} \delta_{\nu}^{\beta}\bigr),\\
I_{14} &=& {\star}R_{\alpha\beta}^{\,\,\,\,\,\,\mu\nu}
           A_{\,\,\,\,\,\,\mu\nu}^{\alpha\beta}
\end{eqnarray}
\end{mathletters}
where $A_{\,\,\,\,\,\,\mu\nu}^{\alpha\beta} = 4
           R_{\rho}^{\alpha}R_{\sigma}^{\rho}R_{\mu}^{\sigma}
           R_{\nu}^{\beta} + 3
           R_{\rho}^{\alpha}R_{\mu}^{\rho}R_{\sigma}^{\beta}
           R_{\nu}^{\sigma}$ and ${\star}R_{\alpha\beta\mu\nu} =
           \frac{1}{2} E_{\alpha\beta\sigma\rho}
           R_{\,\,\,\,\,\,\mu\nu}^{\sigma\rho} =
           \frac{1}{2} E_{\sigma\rho\mu\nu}
           R^{\,\,\,\,\,\,\sigma\rho}_{\alpha\beta}$,
           ${\star}R_{\alpha\beta}^{\,\,\,\,\,\,\mu\nu} =
           \frac{1}{2} E_{\alpha\beta\sigma\rho}
           R^{\sigma\rho\mu\nu}$ with 
           $E_{\alpha\beta\mu\nu} =
           \sqrt{-g} \varepsilon_{\alpha\beta\mu\nu}$ and
           $E^{\alpha\beta\mu\nu} = \frac{-1}{
           \sqrt{-g}}\varepsilon^{\alpha\beta\mu\nu}.$
           Here $\varepsilon_{\alpha\beta\mu\nu}$ is the totally
           antisymmetric Levi-Civita tensor with
           $\varepsilon_{0123} = 1.$
Instead of analysing all 14 invariants mentioned above, one can confine
this study only in 3, namely the scalar curvature $I_1 = R$, 
$I_2 = R_{\mu\nu}^R{\mu\nu}$ and the Kretschmann scalar
$I_3 = R_{\alpha\beta\mu\nu}R^{\alpha\beta\mu\nu}$.
At any regular space-time point, these 3 invariants 
$I-1,\,I_2,\,I_3$ should be finite. Let us rewrite these invariants
in details.

For the Bianchi I metric one finds the scalar curvature (see appendix)
\begin{eqnarray}
I_1 = R = -2 \frac{\ddot \tau -{\dot a}{\dot b} c 
- {\dot b}{\dot c} a - {\dot c}{\dot a} b}{\tau}.
\label{SC}
\end{eqnarray}
Since the Ricci tensor for the Bianchi I metric is diagonal, the 
invariant $I_2 = R_{\mu\nu}R^{\mu\nu} \equiv R_{\mu}^{\nu} R_{\nu}^{\mu}$
is a sum of squares of diagonal components of Ricci tensor, that i.e.,
\begin{equation}
I_2 = \Bigl[\bigl(R_{0}^{0}\bigr)^2  + \bigl(R_{1}^{1}\bigr)^2 +
\bigl(R_{2}^{2}\bigr)^2 + \bigl(R_{3}^{3}\bigr)^2 \Bigr],
\end{equation}
with
\begin{mathletters}
\label{RT1}
\begin{eqnarray}
R_{0}^{0} &=& - \frac{{\ddot a}bc + a{\ddot b}c
+ ab{\ddot c}}{\tau},\\
R_{1}^{1} &=& - \frac{{\ddot a} bc + {\dot a}
{\dot b} c + {\dot a} b {\dot c}}{\tau},\\ 
R_{2}^{2} &=& - \frac{a{\ddot b}c + {\dot a}
{\dot b} c + a{\dot b}{\dot c}}{\tau},\\ 
R_{3}^{3} &=& - \frac{ab{\ddot c} + a{\dot b}
{\dot c} + {\dot a} b{\dot c}}{\tau}. 
\end{eqnarray}
\end{mathletters}
Analogically, for the Kretschmann scalar in this case we have 
$I_3 = R_{\,\,\,\,\,\,\,\alpha\beta}^{\mu\nu}
R_{\,\,\,\,\,\,\,\mu\nu}^{\alpha\beta}$,
a sum of squared components of all nontrivial 
$R_{\,\,\,\,\,\,\,\alpha\beta}^{\mu\nu}$:
\begin{eqnarray}
I_3 &=& 4 \Biggl[ \Bigl(R_{\,\,\,\,\,\,01}^{01}\Bigr)^2 
+ \Bigl(R_{\,\,\,\,\,\,01}^{01}\Bigr)^2 + 
+ \Bigl(R_{\,\,\,\,\,\,02}^{02}\Bigr)^2
+ \Bigl(R_{\,\,\,\,\,\,03}^{03}\Bigr)^2
+ \Bigl(R_{\,\,\,\,\,\,12}^{12}\Bigr)^2
+ \Bigl(R_{\,\,\,\,\,\,23}^{23}\Bigr)^2
+ \Bigl(R_{\,\,\,\,\,\,31}^{31}\Bigr)^2\Biggr] \nonumber\\
&=& \frac{4}{\tau^2}\Bigl[({\ddot a} bc)^2 + (a {\ddot b} c)^2
+ (ab {\ddot c})^2 + ({\dot a}{\dot b} c)^2 + ({\dot a} b {\dot c})^2
+ (a {\dot b}{\dot c})^2\Bigr],\quad \tau = abc.
\label{Kretsch}
\end{eqnarray}
From (\ref{abc}) we have 
\begin{mathletters}
\label{sing}
\begin{eqnarray}
a_i &=& A_i \tau^{1/3} {\rm exp} [(B_i/3) \int \tau^{-1} dt],  
\\
{\dot a}_i &=& \frac{B_i+1}{3}\frac{a_i}{\tau}, \quad (i=1,2,3,)\\
{\ddot a}_i &=& \frac{(B_i+1)(B_i-2)}{9}\frac{a_i}{\tau^2},
\end{eqnarray}
\end{mathletters}
i.e., the metric functions $a, b, c$ and their derivatives are in 
functional dependence with $\tau$. As we see from (\ref{sing}),
at any space-time point, where $\tau = 0$ the invariants $I_1,\,I_2,\,
I_3$ become infinity, hence the space-time becomes singular at this 
point.

\section{Analysis of the results} 

In this section we shall analyze the general results obtained in the
previous section. In the subsections follow we will study the system
with linear and nonlinear scalar fields respectively. 

\subsection{Linear spinor field in B-I universe}

In this subsection we study the linear spinor field in B-I universe.
The reason for getting 
the solution to the self-consistent system of equations for the linear 
spinor and gravitational fields is the necessity of comparing this solution 
with that for the system of equations for the nonlinear spinor and 
gravitational  fields that permits clarifications of the role of nonlinear 
spinor terms in the evolution of the cosmological model in question. 

In this case we get explicit expressions for the components of spinor
field functions and metric functions:
\begin{mathletters}
\label{psil}
\begin{eqnarray} 
\psi_1(t) &=& (C_1/\sqrt{\tau}) {\rm exp}\,[-imt],\\
\psi_2(t) &=& (C_2/\sqrt{\tau}) {\rm exp}\,[-imt],\\
\psi_3(t) &=& (C_3/\sqrt{\tau}) {\rm exp}\,[imt],\\
\psi_4(t) &=& (C_4/\sqrt{\tau}) {\rm exp}\,[imt],
\end{eqnarray} 
\end{mathletters}
with $C_1,\,C_2,\,C_3,\,C_4$ being the integration constants.
On the other hand from (\ref{inv}) we find
\begin{equation}
S = \frac{C_0}{\tau},
\label{Sl}
\end{equation}
where $C_0$ is an integration constant and related to the previous 
ones as $C_0 = C_{1}^{2} + C_{2}^{2} - C_{3}^{2} - C_{4}^{2}.$
For the components of the spin current from (\ref{spincur}) we find
\begin{mathletters}
\label{spincurlin}
\begin{eqnarray}
j^0 &=& \frac{1}{\tau}
\bigl[C_{1}^{2} + C_{2}^{2} + C_{3}^{2} + C_{4}^{2}\bigr],\\
j^1 &=& \frac{2}{a\tau}
\bigl[C_{1} C_{4} + C_{2} C_{3}\bigr] {\rm cos}(2mt),\\
j^2 &=& \frac{2}{b\tau}
\bigl[C_{1} C_{4} - C_{2} C_{3}\bigr] {\rm sin}(2mt),\\
j^3 &=& \frac{2}{c\tau}
\bigl[C_{1} C_{3} - C_{2} C_{4}\bigr] {\rm cos}(2mt),
\end{eqnarray}
\end{mathletters}
whereas, for the projection of spin vectors on the $X$, $Y$ and $Z$
axis we find
\begin{mathletters}
\label{projlin}
\begin{eqnarray}
S^{23,0} &=& \frac{1}{b c\tau}\bigl[C_{1} C_{2} + C_{3} C_{4}\bigr],\\
S^{31,0} &=& 0,\\
S^{12,0} &=& \frac{1}{2ab\tau}
\bigl[C_{1}^{2} - C_{2}^{2} + C_{3}^{2} - C_{4}^{2}\bigr].
\end{eqnarray}
\end{mathletters}
From (\ref{charge}) we find the charge of the system in a volume
${\cal V}$
\begin{equation}
Q = \bigl[C_{1}^{2} + C_{2}^{2} + C_{3}^{2} + C_{4}^{2}\bigr]{\cal V}.
\label{chl}
\end{equation}
Thus we see, the total charge of the system ina finite volume is
always finite.

Let us now determine the function $\tau$. In absence of perfect
fluid for the linear spinor field we have 
\begin{equation}
T_{0}^{0} = m S, \quad T_{1}^{1} = T_{2}^{2} = T_{3}^{3} = 0.
\label{enls}
\end{equation}
Taking (\ref{enls}) into account, for $\tau$ we write 
\begin{equation}
\ddot \tau = M - 3\Lambda \tau
\end{equation}
with the solutions
\begin{equation}
\label{taulin}
\tau = 
\left\{\begin{array}{ccc} 
(1/3\Lambda)\,[M - q_1 {\rm sinh}(\sqrt{-3\Lambda} t)],& \Lambda < 0, \\ 
(1/2) Mt^2 + y_1 t + y_0,& \Lambda = 0, \\ 
(1/3\Lambda)\,[M - q_2 {\rm sin}(\sqrt{3\Lambda} t)],& \Lambda > 0. 
\end{array}\right.
\end{equation}
where $M=\frac{3}{2}\kappa mC_0$ and $y_1, y_0, q_1, q_2$ are the constants.  
Let us now analyze the solutions obtained.  

First we study the case when $\Lambda = 0.$ It can be shown that~
\cite{sahajmp} 
\begin{equation}
y_{1}^{2}- 2My_0\,=\,(X_{1}^{2}+X_1X_3+X_{3}^{2})/3 > 0.
\end{equation} 
This means that the quadratic polynomial $(1/2) M t^2 + y_1 t + y_0 = 0$
possesses real roots, i.e., $\tau(t)$ in case of $\Lambda = 0$
turns into zero at $t=t_{1,2} = -y_1/M \pm \sqrt{(y_1/M)^2 - 2y_0/M}$ 
and the solution obtained is the singular one. 
At $t \to \infty$ in this case we have 
$$\tau(t) \approx \frac{3}{4}\kappa mC_0 t^2, \qquad
a(t) \approx b(t) \approx c(t) \approx t^{2/3}, $$
which leads to the conclusion about the asymptotical 
isotropization of the expansion process for the initially anisotropic 
B-I space. Thus the solution to the self-consistent system of  
equations for the linear spinor and gravitational fields  is  the singular 
one at the initial time. In the initial state of evolution of the field 
system the expansion process of space is  anisotropic, but at 
$t \to \infty$ the isotropization of the expansion process takes place. 
As one sees the components of spin current and projections of spin vector
are singular at space-time points $t_{1,2}$ where $\tau$ vanishes.
A qualitative picture of this case has been given in Fig. 1.

For $\Lambda < 0$ we see that the solution is singular at
$t = t_0 = (1/\sqrt{-3\Lambda}) {\rm arcsinh}(M/q_1)$ and 
the isotropization of the expansion process takes place as 
$t \to \infty.$ Note that the izotropization process in this
case is rather rapid [cf. Fig 2].

For $\Lambda > 0$ we have the oscillatory solutions [cf. Fig. 3]. 
Taking into account that $\tau$ is a non-negative quantity, it can 
be shown that the model has singular solutions at 
$t = (4k + 1) \pi/2\sqrt{3\Lambda}$, $k = 0,1,2,3,....$ with $M = q_2$. 
For $M > q_2$ we have $\tau$ that is always positive definite, i.e., 
the solutions obtained are regular at each space-time point.

\begin{figure}
\hspace{2cm}\epsfig{file=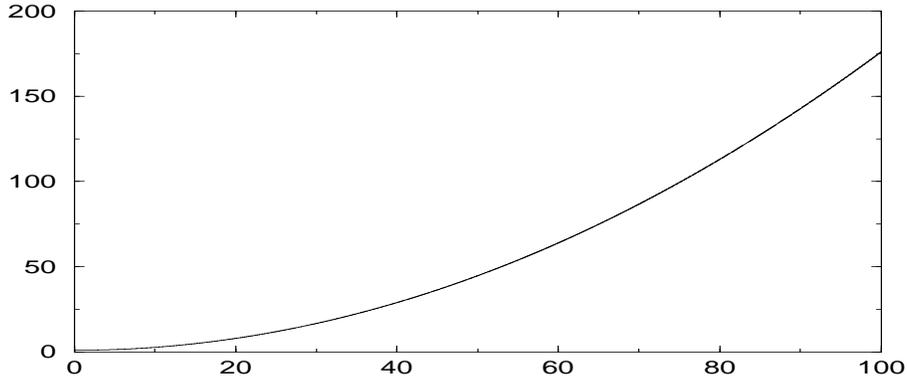,height=5cm,width=12cm,angle=0}
\vspace{.5cm}
\caption{Perpective view of $\tau$ for linear spinor field in 
absence of $\Lambda$ term.}
\end{figure}

\begin{figure}
\hspace{2cm}\epsfig{file=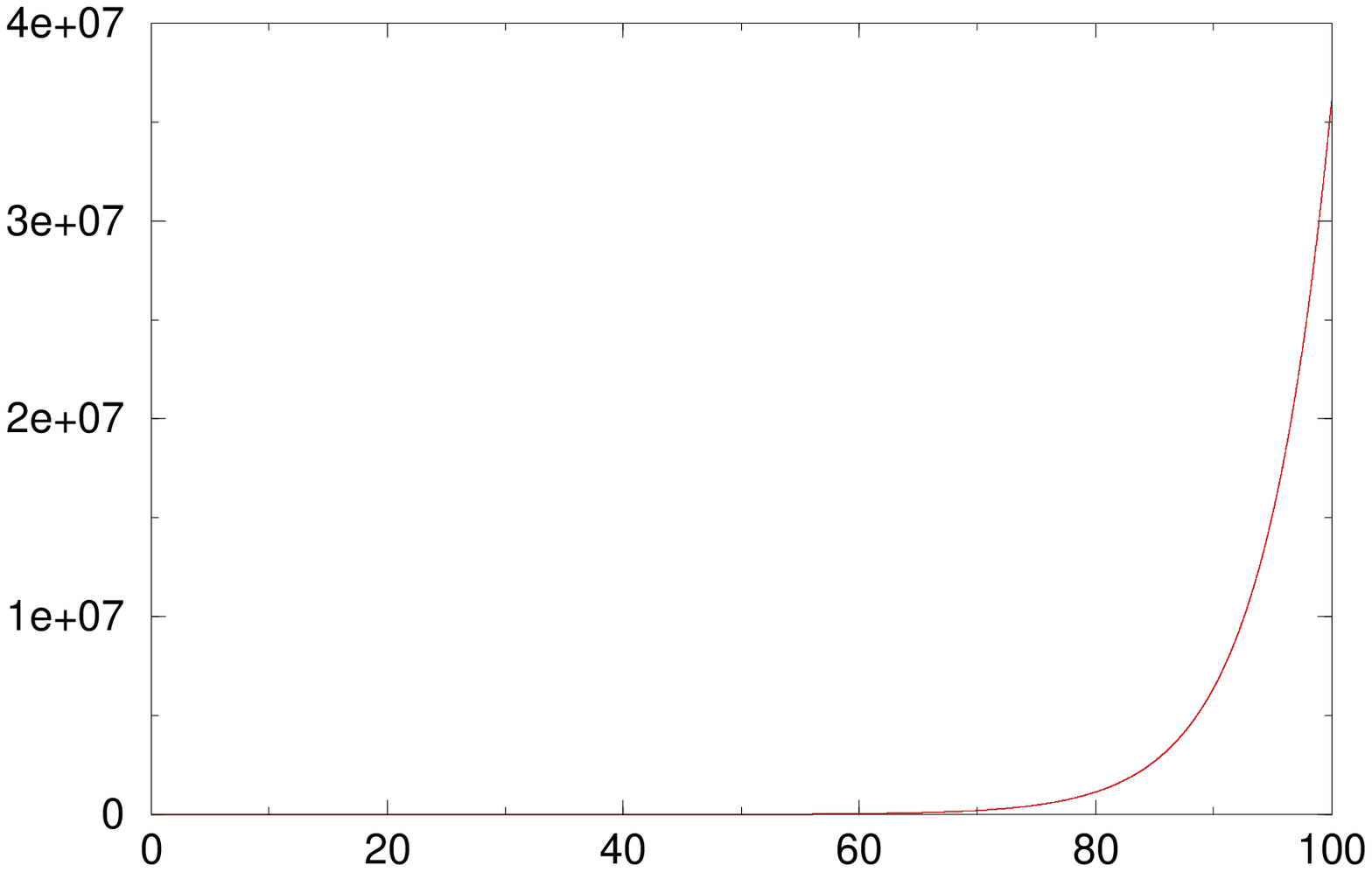,height=5cm,width=12cm,angle=0}
\vspace{.5cm}
\caption{Perpective view of $\tau$ for linear spinor field with 
$\Lambda < 0$.}
\end{figure}

\begin{figure}
\hspace{2cm}\epsfig{file=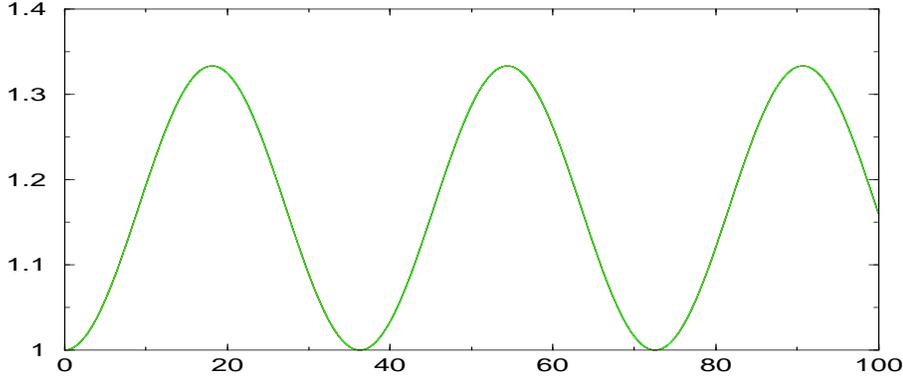,height=5cm,width=12cm,angle=0}
\vspace{.5cm}
\caption{Perpective view of $\tau$ for linear spinor field with 
$\Lambda > 0$.}
\end{figure}

\subsection{Nonlinear spinor field}
Let us now go back to nonlinear case. We consider the following forms 
of nonlinear term: {\bf I.} $L_{\rm N}\,=\,F(I)$; {\bf II.}
$L_{\rm N}\,=\,F(J)$; {\bf III.} $L_{\rm N}\,=\,F(K_{\pm})$ with
$K_{\pm} = I \pm J.$

{\bf I.} Let us consider the case when $L_{\rm N}\,=\,F(I)$.
From (\ref{inv}) we find  in this case we find
\begin{equation}
S = \frac{C_0}{\tau}, \quad C_0= {\rm const.}
\label{SNL}
\end{equation}
Note that in this case we denote the constants in the same way as we did 
it for linear case, but the constants in these cases are not necessarily
identical. Spinor field equations in this case read
\begin{mathletters}
\label{VI}
\begin{eqnarray} 
\dot{V}_{1} + i (m - {\cD}) V_{1} &=& 0, \\
\dot{V}_{2} + i (m - {\cD}) V_{2} &=& 0, \\
\dot{V}_{3} - i (m - {\cD}) V_{3} &=& 0, \\
\dot{V}_{4} - i (m - {\cD}) V_{4} &=& 0. 
\end{eqnarray} 
\end{mathletters}
As in the considered case when $L_N\,=\,F$ depends only on $S$, 
from (\ref{SNL}) it follows that $F(I)$ and ${\cD}$ are functions of 
$\tau$ only. Taking this fact into account, 
we get explicit expressions for the components of spinor
field functions
\begin{mathletters}
\label{psinl}
\begin{eqnarray} 
\psi_1(t) &=& (C_1/\sqrt{\tau}) {\rm exp}\,[-i\int(m - {\cD}) dt],\\
\psi_2(t) &=& (C_2/\sqrt{\tau}) {\rm exp}\,[-i\int(m - {\cD}) dt],\\
\psi_3(t) &=& (C_3/\sqrt{\tau}) {\rm exp}\,[i\int(m - {\cD}) dt],\\
\psi_4(t) &=& (C_4/\sqrt{\tau}) {\rm exp}\,[i\int(m - {\cD}) dt],
\end{eqnarray} 
\end{mathletters}
with $C_1,\,C_2,\,C_3,\,C_4$ being the integration constants and
are related to $C_0$ as 
$C_0 = C_{1}^{2} + C_{2}^{2} - C_{3}^{2} - C_{4}^{2}.$
For the components of the spin current from (\ref{spincur}) we find
\begin{mathletters}
\label{spincurnlf}
\begin{eqnarray}
j^0 &=& \frac{1}{\tau}
\bigl[C_{1}^{2} + C_{2}^{2} + C_{3}^{2} + C_{4}^{2}\bigr],\\
j^1 &=& \frac{2}{a\tau}
\bigl[C_{1} C_{4} + C_{2} C_{3}\bigr] {\rm cos}[2\int(m - {\cD}) dt],\\
j^2 &=& \frac{2}{b\tau}
\bigl[C_{1} C_{4} - C_{2} C_{3}\bigr] {\rm sin}[2\int(m - {\cD}) dt],\\
j^3 &=& \frac{2}{c\tau}
\bigl[C_{1} C_{3} - C_{2} C_{4}\bigr] {\rm cos}[2\int(m - {\cD}) dt],
\end{eqnarray}
\end{mathletters}
whereas, for the projection of spin vectors on the $X$, $Y$ and $Z$
axis we find
\begin{mathletters}
\label{projnlf}
\begin{eqnarray}
S^{23,0} &=& \frac{1}{b c\tau}\bigl[C_{1} C_{2} + C_{3} C_{4}\bigr],\\
S^{31,0} &=& 0,\\
S^{12,0} &=& \frac{1}{2ab\tau}
\bigl[C_{1}^{2} - C_{2}^{2} + C_{3}^{2} - C_{4}^{2}\bigr].
\end{eqnarray}
\end{mathletters}

We now study the equation for $\tau$ in details choosing
the nonlinear spinor term as $F(I) =\lambda I^{(n/2)}=\lambda S^n$ 
with $\lambda$ being the coupling constant and $n>1$. 
In this case for $\tau$ one gets 
\begin{equation}
\ddot \tau = (3/2)\kappa \bigl[m C_0 + \lambda (n-2) 
C_{0}^{n}/\tau^{n-1}\bigr] - 3 \Lambda \tau.
\label{taui}  
\end{equation}
The first integral of the foregoing equation takes form
\begin{equation}
\dot \tau^2 = 3\kappa \bigl[m C_0 \tau - \lambda 
C_{0}^{n}/\tau^{n-2} + g^2\bigr] - 3 \Lambda \tau^2. 
\label{fii}
\end{equation} 
Here $g^2$ is the integration constant that is positive defiend
and connected with the constants $X_i$ as
$g^2 =  (X_1^2 + X_1 X_3 + X_3^2)/9\kappa\,C_0$~\cite{sahajmp}.
The sign $C_0$  is determined by the positivity 
of the energy-density $T_{0}^{0}$ of linear spinor field, i.e.,
\begin{equation}
T_{0}^{0} = m C_0/\tau > 0.
\label{enls1}
\end{equation}   
It is obvious from (\ref{enls1}) that $C_0 >0.$ Now one can write the
solution to the equation (\ref{fii}) in quadratures:
\begin{equation}
\int \frac{\tau^{(n-2)/2}d\tau}{\sqrt{\kappa[m C_0 \tau^{n-1} +g^2 
\tau^{n-2} - \lambda C_{0}^{n}] - \Lambda \tau^n}}= \sqrt{3}\,t
\label{qudri} 
\end{equation} 
The constant of integration in (\ref{qudri}) has been taken to be zero, 
as it only gives the shift of the initial time. Let us study the 
properties of solution obtained for different choice of $n, \lambda$
and $\Lambda$. Firts we study the case with $\Lambda = 0.$

For $n > 2$ from (\ref{qudri}) one gets 
\begin{equation}
\tau(t)\mid_{t \to \infty} \approx (3/4) \kappa mC_0t^2.
\label{ng2}
\end{equation}
It leads to the conclusion about isotropization of the expansion process 
of the B-I space-time. It should be remarked that the isotropization takes 
place if and only if the spinor field equation contains the massive term  
[cf. the parameter $m$ in (\ref{qudri})]. This is not the case for a 
massless spinor field, since from (\ref{qudri}) we get 
\begin{equation}
\tau(t)\mid_{t \to \infty} \approx \sqrt{3\kappa C_0 g^2}\,t.
\label{reg}
\end{equation} 
Substituting (\ref{reg}) into (\ref{abc}) one comes to the conclusion 
that the functions $a(t), b(t)$, and $c(t)$ are different. 

Let us consider the properties of solutions to Eq. (\ref{taui}) when 
$t \to 0.$ For $\lambda<0$ from (\ref{qudri}) we get 
\begin{equation}
\tau(t)= \bigl[(3/4) n^2 \kappa |\lambda| 
C_{0}^{n}\bigr]^{1/n}t^{2/n} \to 0, 
\end{equation} 
i.e. solutions are singular. For $\lambda>0,$ from (\ref{qudri}) 
it follows that $\tau=0$ cannot be reached for any value of $t$ 
as in this case when the denominator of the integrand in (\ref{qudri}) 
becomes imaginary. It means that for $\lambda>0$ there exist regular 
solutions to the previous system of equations~\cite{sahactp1}. 
The absence of the initial singularity in the 
considered cosmological solution appears to be consistent with the 
violation for $\lambda>0$ of the dominant energy condition in the 
Hawking-Penrose theorem~\cite{hawking} which reads as follows:

{\bf THEOREM.} {\it A space-time ${\cal M}$ cannot be causally,
geodesically complete if the GTR equations hold and if the following
conditions are fulfilled:}

1. The space-time ${\cal M}$ does not contain closed time-like lines.

2. The conditions (dominant energy condition)
\begin{mathletters}
\label{dec1}
\begin{eqnarray}
T_{00} + T_{11} + T_{22} + T_{33} &\ge& 0,\\
T_{00} + T_{11} &\ge& 0, \\
T_{00} + T_{22} &\ge& 0, \\
T_{00} + T_{33} &\ge& 0, 
\end{eqnarray}
\end{mathletters}
on the equations of state are fulfilled, where $T_{00}$ is the energy 
density and $T_{11},\,T_{22},$ and $T_{33}$ are three principal values of
pressure tensor.

3. On each time-like or null geodesic, there is at least one point for
which
\begin{equation}
K_{[a}R_{b]cd[e}K_{f]}K^c K^d \ne 0,
\label{c3}
\end{equation}
where $K_a$ is the tangent to the curve at the given point and where
the brackets on the subscripts imply antisymmetrization.

4. The space-time ${\cal M}$ contains either $(a)$ a point $P$
such that all diverging rays from this point begin to converge
if one traces them back into the past, or $(b)$ a compact space-like
hypersurface. 

To prove that in the case considered the dominant energy condition 
violates, we rewrite (\ref{dec1}) in the following form:
\begin{mathletters}
\label{dec2}
\begin{eqnarray}
T_{0}^{0} &\ge& T_{1}^{1} a^2 + T_{2}^{2} b^2 + T_{3}^{3} c^2,\\
T_{0}^{0} &\ge& T_{1}^{1} a^2, \\
T_{0}^{0} &\ge& T_{2}^{2} b^2, \\
T_{0}^{0} &\ge& T_{3}^{3} c^2.
\end{eqnarray}
\end{mathletters}
Let us go back to the energy density of spinor field. From 
\begin{equation}
T_{0}^{0} = \frac{mC_0}{\tau} - \frac{\lambda C_0^n}{\tau^n}
\end{equation} 
follows that at
\begin{equation}
\tau^{n-1} < \frac{\lambda C_0^{n-1}}{m}
\end{equation}
the energy density of spinor field becomes negative. 
On the other hand we have
\begin{equation}
T_{1}^{1} = T_{2}^{2} = T_{3}^{3} = \frac{\lambda (n-1) C_0^n}{\tau^n} > 0
\end{equation}
for any non-negative value of $\tau$. Thus, we see all four conditions
in (\ref{dec2}) violate, i.e., 
the absence of initial singularity in the considered cosmological 
solution appears to be consistent with the violation of the dominant 
energy condition in the Hawking-Penrose theorem. 

Let us consider the Heisenberg-Ivanenko equation~\cite{ddiv} setting
$n = 2$ in (\ref{taui}). 
In this case  the equation  for $\tau(t)$ does not contain 
the nonlinear term and its solution coincides with that of the linear 
one. The spinor field functions in this case are written as follows:  
\begin{mathletters}
\begin{eqnarray}
V_1 &=& \frac{C_1}{\sqrt{\tau}}\,e^{-imt}Z^{4i\lambda C_0/B}, \\
V_2 &=& \frac{C_2}{\sqrt{\tau}}\,e^{-imt}Z^{4i\lambda C_0/B}, \\
V_3 &=& \frac{C_3}{\sqrt{\tau}}\,e^{imt}Z^{-4i\lambda C_0/B},\\ 
V_4 &=& \frac{C_4}{\sqrt{\tau}}\,e^{imt}Z^{-4i\lambda C_0/B}, 
\end{eqnarray}
\end{mathletters}
where $Z = \frac{(t-t_1)}{(t-t_2)},\,\, B = M(t_1 -t_2),$ 
and $t_{1,2} = -y_1/M \pm \sqrt{(y_1/M)^2 - 2y_0/M}$\, are the 
roots  of the quadratic equation \, $Mt^2+2y_1t+2y_0 = 0.$ 
As in the linear case, the obtained solution is singular at initial time 
and asymptotically isotropic as $t \to \infty$.

We now study the properties of solutions to Eq. (\ref{taui}) for 
$1<n<2.$ In this case it is convenient to present the solution 
(\ref{qudri}) in the form
\begin{equation}
\int \frac{d \tau}{\sqrt{m\tau -\lambda \tau^{2-n} 
C_{0}^{n-1}+g^2}}=\sqrt{3\kappa C_0}\,t 
\label{1n2}
\end{equation} 
As $t \to \infty$, from (\ref{1n2}) we get the equality (\ref{ng2}), 
leading to the isotropization of the expansion process. If $m=0$ and 
$\lambda>0,$ \quad $\tau(t)$ lies on the interval 
$$0 \le \tau(t) \le \bigl(g^2/\lambda C_{0}^{n-1}\bigr)^{1/(2-n)}.$$
If $m=0$ and $\lambda<0,$ the relation (\ref{1n2}) at $t \to \infty$ 
leads to the equality  
\begin{equation}
\tau(t) \approx \bigl[(3/4)n^2 \kappa |\lambda| C_{0}^{n} 
\bigr]^{1/n}t^{2/n}.
\label{ll0}
\end{equation}  
Substituting (\ref{ll0}) into (\ref{abc}) and taking into account that 
at $t \to \infty$
$$ \int \frac{dt}{\tau} \approx \frac{n(3\kappa |\lambda| 
n^2C_{0}^{n})^{1/n}}{(n-2)2^{2/n}}t^{-2/n+1} \to 0 $$
due to $-2/n+1 < 0,$ we obtain 
\begin{equation}
a(t) \sim b(t) \sim c(t) \sim 
[\tau(t)]^{1/3} \sim t^{2/3n} \to \infty.  
\label{abcll0}
\end{equation} 
This means that the solution obtained tends to the isotropic one. In this 
case the isotropization is provided not by the massive parameter, but by 
the degree $n$ in the term $L_N = \lambda S^n.$ Equation (\ref{1n2}) 
implies 
\begin{equation}
\tau(t)\mid_{t \to 0} \approx \sqrt{3\kappa C_0 g^2}\,t \to 0,
\label{t0}
\end{equation} 
which means the solution obtained is initially singular. Thus for $1<n<2$ 
there exist only singular solutions at initial time. At $t \to \infty$ 
the isotropization of  the expansion process of the B-I space 
takes place both for $m\not= 0$ and for $m=0.$  

Finally, let us study the properties of   the solution to the equation 
(\ref{taui}) for $0<n<1.$ In this case  we use the solution in the form 
(\ref{1n2}). Since now $2-n>1,$ then with the increasing of $\tau(t)$ 
in the denominator of the integrand in (\ref{1n2}) the second term 
$\lambda \tau^{2-n} C_{0}^{n-1}$ increases faster than the first one.  
Therefore the solution describing the space expansion can be possible 
only for $\lambda<0.$  In this case at $t\to \infty$, for $m=0$ as well 
as for $m\not= 0,$ one can get the asymptotic representation (\ref{ll0}) 
of the solution. This solution, as for the choice $1<n<2,$ provides 
asymptotically isotropic expansion of the B-I space-time. For $t \to 0$ 
in this case we shall get only the singular solution of the form (\ref{t0}). 

For a nonzero $\Lambda$ term we study the following situations
depending on the sign of $\Lambda$ and $\lambda$. 

{\bf case 1.} $\Lambda = - \epsilon^2 < 0$, $\lambda > 0$. In this case for 
$n > 2$ and $t \to \infty$ we find
\begin{equation}
\tau (t) \approx e^{\sqrt{3} \epsilon t}
\label{c1}
\end{equation}  
Thus we see that the asymptotic behavior of $\tau$ does not depend on
$n$ and defined by $\Lambda$ - term. From (\ref{abc}) it is obvious that
the asymptotic isotropization takes place.

From (\ref{qudri}) it also follows that $\tau$ cannot be zero
at any moment, since the intigrant turns out to be imaginary as $\tau$
approaches to zero. Thus the solution obtained is a nonsingular one
thanks to the nonlinear term in the Dirac equation and asymptotically
isotropic. As it has been noted earlier, the absence of initial 
singularity in the considered cosmological model results the violation
of the dominant energy condition.

{\bf case 2.} $\Lambda > 0$ and $\lambda > 0$. For $n > 2$ 
(\ref{qudri}) admits only nonsingular oscillating solutions, since
$\tau > 0$ and bound from above. Consider the case with $n = 4$ and
for simplicity set $m = 0$. Then from (\ref{qudri}) one gets
\begin{equation}
\tau(t) = \frac{1}{\sqrt{2 \Lambda}} \Bigl[\kappa C_0 \tau_0
+ \sqrt{\kappa^2 C_0^2 \tau_0^2 + 4 \Lambda \lambda C_0^4}\, 
{\rm sin} 2 \sqrt{3 \Lambda} t \Bigr]^{1/2}.
\end{equation}

For a massive spinor field with $\Lambda > 0$ and $\lambda > 0$
and $n = 10$ a perspective view of $\tau$ is shown in FIG. 4.
The period for the massive field is greater than that for the
massless one. As it occurs, the order of nonlinearity $(n)$ 
has a direct effect on the period (the more in $n$ the less 
is the period)

\vspace*{.5cm}
\begin{figure}
\hspace{.1cm}\epsfig{file=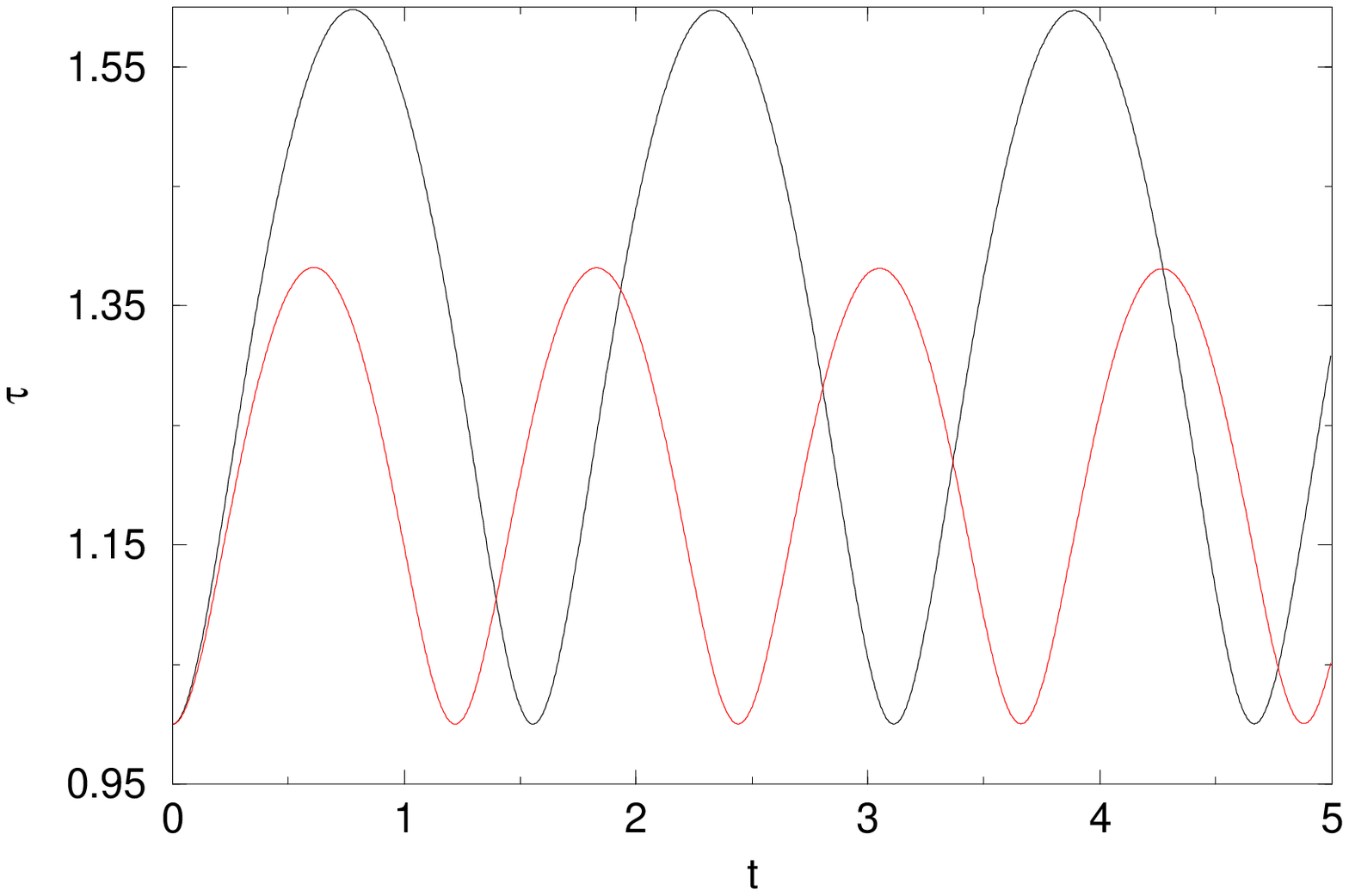,height=6cm,width=12cm,angle=0}
\vspace{.5cm}
\caption{Perspective view of $\tau$ showing the initially
nonsingular and oscillating behavior of the solutions. The 
continuous and dash lines correspond to the massive and massless
spinor field, respectively.}
\end{figure}

{\bf case 3.} $\Lambda < 0$ and $\lambda < 0$. The solution is
singular at initial moment, that is
\begin{equation}
\lim\limits_{t \to 0} \tau \approx [\sqrt{-3 \lambda n^2 C_0^n/4}t]^{2/n}
\end{equation} 
and at $t \to \infty$ asymptotic isotropization takes place since
\begin{equation}
\lim\limits_{t \to \infty} \tau \approx e^{\sqrt{3 \Lambda} t}.
\end{equation} 

{\bf case 4.} $\Lambda > 0$ and $\lambda < 0$. Solution is initially
singular as 
\begin{equation}
\lim\limits_{t \to 0} \tau \approx [\sqrt{-3 \lambda n^2 C_0^n/4}t]^{2/n}
\end{equation} 
and bound from the above, i.e., oscillating, since
\begin{equation}
\lim\limits_{t \to \infty} \tau \approx {\rm sin} \sqrt{3 \Lambda} t.
\end{equation}

{\bf II.} We study the system when $L_N\,=\,F(J)$, which means in the case 
considered ${\cal D}\,=\,0$. Let us note that, in the unified 
nonlinear spinor theory of Heisenberg, the massive term remains 
absent, and according to Heisenberg, the particle mass should be 
obtained as a result of quantization of spinor prematter~
\cite{massless}. In the nonlinear generalization of classical field 
equations, the massive term does not possess the significance that 
it possesses in the linear one, as it by no means defines total 
energy (or mass) of the nonlinear field system. Thus without losing 
the generality we can consider massless spinor field putting $m\,=\,0.$ 
Then from (\ref{inv}) one gets
\begin{equation}
P(t) = \frac{D_0}{\tau}, \quad D_0 = {\rm const.}
\label{pinv}
\end{equation}
The system of spinor field equations in this case reads
\begin{mathletters}
\label{VJ}
\begin{eqnarray} 
\dot{V}_{1} - {\cG} V_{3} &=& 0, \\
\dot{V}_{2} - {\cG} V_{4} &=& 0, \\
\dot{V}_{3} + {\cG} V_{1} &=& 0, \\
\dot{V}_{4} + {\cG} V_{2} &=& 0. 
\end{eqnarray} 
\end{mathletters}
Defining $U(\sigma) = V (t)$, where $\sigma = \int {\cG} dt$, we 
rewrite (\ref{VJ}) as
\begin{mathletters}
\label{UJ}
\begin{eqnarray} 
U_{1}^{\prime} - U_{3} &=& 0, \\
U_{2}^{\prime} - U_{4} &=& 0, \\
U_{3}^{\prime} + U_{1} &=& 0, \\
U_{4}^{\prime} + U_{2} &=& 0, 
\end{eqnarray} 
\end{mathletters}
where prime ($^\prime$) denote differentiation with respect to $\sigma$.
Differentiating the first equation of system (\ref{UJ}) and taking into 
account the third one we get 
\begin{equation}
U_{1}^{\prime \prime} +U_{1} =\,0,
\end{equation}
which leads to the solution
\begin{eqnarray}
U_1 &=& D_1 e^{i \sigma} + iD_3 e^{-i \sigma},\nonumber\\
U_3 &=&  i D_1 e^{i \sigma} + D_3 e^{-i \sigma}.\nonumber
\end{eqnarray}
Analogically for $U_2$ and $U_4$ one gets
\begin{eqnarray}
U_2 &=& D_2 e^{i \sigma} + iD_4 e^{-i \sigma},\nonumber\\
U_4 &=& i D_2 e^{i \sigma} + D_4 e^{-i \sigma}, \nonumber
\end{eqnarray}
where $D_i$ are the constants of integration.
Finally, we can write
\begin{mathletters}
\label{psij}
\begin{eqnarray}
\psi_1 &=&\frac{1}{\sqrt{\tau}} \bigl(D_1 e^{i \sigma} + 
iD_3 e^{-i\sigma}\bigr), \\
\psi_2 &=&\frac{1}{\sqrt{\tau}} \bigl(D_2 e^{i \sigma} + 
iD_4 e^{-i\sigma}\bigr),  \\
\psi_3 &=&\frac{1}{\sqrt{\tau}} \bigl(iD_1 e^{i \sigma} + 
D_3 e^{-i \sigma}\bigr), \\
\psi_4 &=&\frac{1}{\sqrt{\tau}} \bigl(iD_2 e^{i \sigma} + 
D_4 e^{-i\sigma}\bigr).
\end{eqnarray} 
\end{mathletters}
Putting (\ref{psij}) into the expressions (\ref{pinv}) one comes
to 
$$D_0=2\,(D_{1}^{2} + D_{2}^{2} - D_{3}^{2} -D_{4}^{2}).$$
For the components of the spin current from (\ref{spincur}) we find
\begin{mathletters}
\label{spincurnlj}
\begin{eqnarray}
j^0 &=& \frac{2}{\tau}
\bigl[D_{1}^{2} + D_{2}^{2} + D_{3}^{2} + D_{4}^{2}\bigr],\\
j^1 &=& \frac{4}{a\tau}
\bigl[D_{2} D_{3} + D_{1} D_{4}\bigr] {\rm cos}[2\int {\cG}_1 dt],\\
j^2 &=& \frac{4}{b\tau}
\bigl[D_{2} D_{3} - D_{1} D_{4}\bigr] {\rm sin}[2\int {\cG}_1 dt],\\
j^3 &=& \frac{4}{c\tau}
\bigl[D_{1} D_{3} - D_{2} D_{4}\bigr] {\rm cos}[2\int {\cG}_1) dt],
\end{eqnarray}
\end{mathletters}
whereas, for the projection of spin vectors on the $X$, $Y$ and $Z$
axis we find
\begin{mathletters}
\label{projnlj}
\begin{eqnarray}
S^{23,0} &=& \frac{2}{b c\tau}\bigl[D_{1} D_{2} + D_{3} D_{4}\bigr],\\
S^{31,0} &=& 0,\\
S^{12,0} &=& \frac{1}{2ab\tau}
\bigl[D_{1}^{2} - D_{2}^{2} + D_{3}^{2} - D_{4}^{2}\bigr].
\end{eqnarray}
\end{mathletters}

Let us now estimate $\tau$ using the equation
\begin{equation}
\ddot{\tau}/\tau\,=\,3 \kappa \,\lambda (n - 1) P^{2n},
\label{pdd}
\end{equation}
where we chose $L_N\,=\,\lambda P^{2n}$. Putting the value of $P$ into
(\ref{pdd}) and integrating one gets
\begin{equation}
\dot{\tau}^2  = - 3\kappa\,\lambda D_{0}^{2n} \tau^{2 - 2n} + y^2,
\label{pfd}
\end{equation}
where $y^2$ is the integration constant having the form
$y^2 = (X_1^2 + X_1 X_3 + X_3^2)/3 > 0$. 
The solution to the equation (\ref{pfd}) in quadrature reads
\begin{equation}
\int\,\frac{d\tau}{\sqrt{- 3 \kappa\lambda D_{0}^{2n}\tau^{2 - 2n} + y^2}} 
= t.
\label{pqud}
\end{equation}
Let us now analyze the solution obtained here. As one can see
the case $n = 1$ is the linear one.  In case of $\lambda < 0$ for
$n > 1$ i.e. $2 - 2n < 0$, we get
$$ \tau(t)\mid_{t \to 0} \approx [(\sqrt{3 \kappa|\lambda|}  
D_{0}^{n}n) t]^{1/n},$$
and
$$ \tau\mid_{t \to \infty} \approx \sqrt{3\kappa y^2} \,t. $$
This means that for the term $L_N$ considered with $\lambda < 0$ and
$n > 1$, the solution is initially singular and the space-time is
anisotropic at $t \to \infty.$ Let us now study it for $n < 1$. In 
this case we obtain
$$ \tau\mid_{t \to 0} \approx \sqrt{3\kappa y^2}\, t $$
and
$$ \tau\mid_{t \to \infty} \approx 
[(\sqrt{3\kappa|\lambda|} D_{0}^{n} n) t]^{1/n}.$$
The solution is initially singular as in the previous case, but as far as
$ 1/n > 1$, it provides an asymptotically isotropic expansion of B-I 
space-time. The analysis for $\Lambda \ne 0$ completely coincides with
those for $F = \lambda S^n$ with $m = 0$.

\vskip 3mm
{\bf III.} In this case we study $L_N\,=\,F(I,\,J)$. Choosing 
\begin{equation}
L_N\,=\,F(K_{\pm}), \quad K_{+} = I + J = I_v = -I_A, \quad
K_{-} = I - J = I_T,
\end{equation}
in the case of massless NLSF we find
$$
{\cal D}\,=\,2 S F_{K_{\pm}}, \quad
{\cal G}\,=\, \pm 2 P F_{K_{\pm}}, \quad F_{K_{\pm}} = dF/dK_{\pm}.
$$
Putting them into (\ref{inv}) we find
\begin{equation}
S_{0}^{2} \pm P_{0}^{2} = D_{\pm}.
\end{equation}
Choosing $F = \lambda K_{\pm}^{n}$ from (\ref{dtau}) we get
\begin{equation}
\ddot{\tau}\,=\,3 \kappa \lambda (n - 1)\,D_{\pm}^{n}\,\tau^{1-2n},
\end{equation}
with the solution
\begin{equation}
\int\,\frac{\tau^{n-1} d \tau}{\sqrt{g^2\tau^{2n - 2} - 3\kappa\lambda 
D_{\pm}^{n}}}\,=\, t,
\end{equation}
where $g^2 = (X_1^2 + X_1 X_3 + X_3^2)/3.$ 
Let us study the case with $\lambda < 0$. For $n < 1$ from (3.33) one gets
\begin{equation}
\tau (t)\mid_{t \to 0} \approx g t \to 0,
\end{equation}
i.e., the solutions are initially singular, and  
\begin{equation}
\tau (t)\mid_{t \to \infty} \approx 
[\sqrt{(3\kappa |\lambda| D_{\pm}^{n})}t]^{1/n},
\end{equation}
which means that the anisotropy disappears as the Universe expands.
In the case of $n > 1$ we get 
$$ \tau (t)\mid_{t \to 0} \approx t^{1/n} \to 0, $$
and
$$ \tau (t)\mid_{t \to \infty} \approx gt, $$
i.e., the solutions are initially singular and the metric functions 
$a(t), b(t)$, and $c(t)$ are different at $ t \to \infty$, i.e.,
the isotropization process remains absent. For $\lambda > 0$ we get 
that the solutions are initially regular, but it violates the
dominant energy condition in the Hawking-Penrose theorem~\cite{hawking}.
Note that one comes to the analogical conclusion choosing 
$L_N\,=\,\lambda S^{2n}P^{2n}.$

\subsection{Analysis of the results obtained when the B-I 
Universe is filled with perfect fluid}

Let us now analyze the system filled with perfect fluid.
In absence of other matter, i.e., spinor field,
in this case from (\ref{dtau}) we find
\begin{equation}
\frac{\ddot \tau} = \frac{3 \kappa}{2}\frac{(1 -\zeta)
\varepsilon_0}{\tau^\zeta},
\end{equation}
with the first integral
\begin{equation}
\dot \tau = \sqrt{3\kappa \ve_0 \tau^{(1 - \zeta)} + C},
\label{pffi}
\end{equation}
where $C$ is an integration constant. From (\ref{pffi}) one estimates
\begin{mathletters}
\label{pfr}
\begin{eqnarray}
\tau &\propto& t^2, \quad {\rm for}\quad \zeta = 0,\quad {\rm (dust)},\\ 
\tau &\propto& t^{3/2}, \quad {\rm for}\quad \zeta = 1/3,\quad 
{\rm (radiation)},\\ 
\tau &\propto& t^{6/5}, \quad {\rm for}\quad \zeta = 2/3,\quad {\rm (hard\,\,\,
universe)},\\ 
\tau &\propto& t, \quad {\rm for}\quad \zeta = 1,\quad {\rm (stiff\,\,\,matter)}. 
\end{eqnarray}
\end{mathletters}
A perspective view of these solutions are given in Fig. 5.

\begin{figure}
\hspace{2cm}\epsfig{file=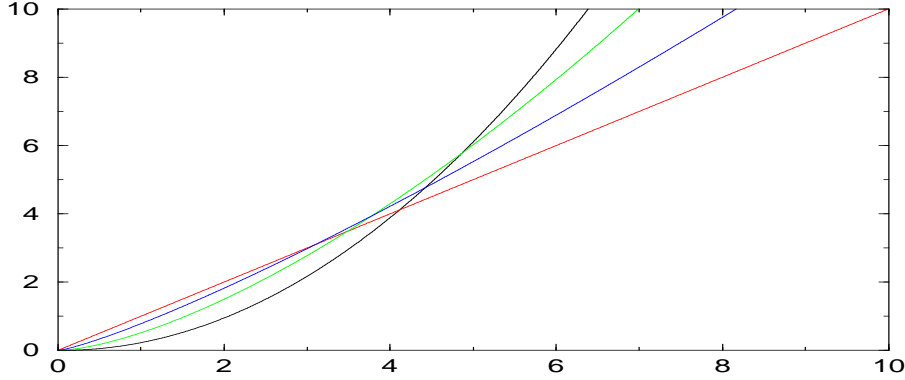,height=5cm,width=12cm,angle=0}
\vspace{.5cm}
\caption{Perpective view of $\tau$ when B-I universe is filled with
perfect fluid only. The lines from left to right at the upper corner
correspond to dust ($\zeta = 0$), radiation ($\zeta = 1/3$), 
hard universe ($\zeta = 2/3$) and stiff matter ($\zeta = 1$), respectively.}
\end{figure}

Let us now consider the system as a whole with the
nonlinear term being $L_{\rm N} = \lambda S^n$. 
In this case we get
\begin{equation}
\int\,\frac{d \tau}{\sqrt{m C_0 \tau - \lambda C_{0}^{n}/\tau^{(n-2)} +
\varepsilon_0 \tau^{(1 - \xi)} + g^2}} = \pm \sqrt{3 \kappa} t.
\label{pfI}
\end{equation}
As one can see in the case of dust $(\xi = 0)$  the fluid term can be 
combined with the massive one, whereas in the case of stiff matter 
$(\xi = 1)$ it mixes up with the constant. Analyzing the equation 
(~\ref{pfI}) one concludes that in presence of spinor field 
perfect fluid plays a secondary role in the evolution of B-I universe.

\section{Conclusion}
Within the framework of the simplest nonlinear model of spinor field 
it has been shown that the $\Lambda$ term plays very important role 
in Bianchi-I cosmology. In particular, it invokes oscillations in the
model which is not the case when $\Lambda$ term remain absent. 
It should be noted that regularity of the solutions obtained
by virtue of $\Lambda$ term, specially for the linear spinor field
does not violate dominant energy condition, while this is not the case 
when regular solutions are attained by means of nonlinear term.
Growing interest in studying the role $\Lambda$ term by present day 
physicists of various discipline witnesses its fundamental value.
For details on time depending $\Lambda$ term one may consult~\cite{sahal}
and references therein.

\end{document}